\documentclass[submission, Phys]{SciPost}

\usepackage{amsmath}
\usepackage{graphicx,multirow,subcaption}
\numberwithin{equation}{section}

\newcommand{\dr}[1]{\widetilde{#1}}

\newcommand{\e}{\text{e}}
\newcommand{\tr}{\text{tr}}

\begin{document}

\begin{center}{\Large \textbf{
The spin Drude weight of the XXZ chain\\*[0.1cm] and generalized hydrodynamics
}}\end{center}

\begin{center}
A. Urichuk\textsuperscript{1,2},
Y. Oez\textsuperscript{1},
A. Kl\"umper\textsuperscript{1},
J. Sirker\textsuperscript{2*}
\end{center}

\begin{center}
{\bf 1} Fakult\"at f\"ur Mathematik und Naturwissenschaften,
Bergische Universit\"at Wuppertal, 42097 Wuppertal, Germany
\\
{\bf 2} Department of Physics and Astronomy, University of Manitoba, Winnipeg R3T 2N2, Canada
\\
* sirker@physics.umanitoba.ca
\end{center}

\begin{center}
\today
\end{center}


\section*{Abstract}
{\bf Based on a generalized free energy we derive exact thermodynamic
Bethe ansatz formulas for the expectation value of the spin current,
the spin current-charge, charge-charge correlators, and consequently
the Drude weight. These formulas agree with recent conjectures within
the generalized hydrodynamics formalism. They follow, however,
directly from a proper treatment of the operator expression of the
spin current. The result for the Drude weight is identical to the one
obtained 20 years ago based on the Kohn formula and TBA. We
numerically evaluate the Drude weight for anisotropies
$\Delta=\cos(\gamma)$ with $\gamma = n\pi/m$, $n\leq m$ integer and
coprime. We prove, furthermore, that the high-temperature asymptotics
for general $\gamma=\pi n/m$---obtained by analysis of the quantum
transfer matrix eigenvalues---agrees with the bound which has been
obtained by the construction of quasi-local charges.}

\vspace{10pt}
\noindent\rule{\textwidth}{1pt}
\tableofcontents\thispagestyle{fancy}
\noindent\rule{\textwidth}{1pt}
\vspace{10pt}

\section{Introduction}
\label{sec:intro}
The Hamiltonian of the XXZ chain is given by
\begin{equation}
\label{Ham}
H=J\sum_{l=1}^N \left(\sigma^x_l\sigma^x_{l+1} +\sigma^y_l\sigma^y_{l+1} +\Delta\sigma^z_l\sigma^z_{l+1}\right)-2h\sum_{l=1}^N\sigma^z_l \, ,
\end{equation}
where $\sigma^{x,y,z}$ are Pauli matrices, $\Delta=\cos(\gamma)$ is
the anisotropy, $h$ the applied magnetic field, and we use periodic
boundary conditions. The XXZ chain is a Bethe ansatz (BA) integrable
model and a family of commuting transfer matrices,
$[T(\theta),T(\theta')]=0$, exists with $\theta$ being the spectral
parameter. The logarithm of the transfer matrix is the generating
function for an infinite set of conserved charges
\begin{equation}
\label{Charge}
Q_n = \frac{d^n}{d\theta^n} \ln T(\theta)\bigg|_{\theta =0}.
\end{equation}
In particular, $Q_1\propto H$ and $Q_2\propto J_E$ where $J_E$ is the
energy current operator. Based on the infinite number of conservation
laws, one might expect that the XXZ chain shows purely ballistic
transport. This is indeed the case for thermal transport
because the energy current $J_E$ is itself a conserved charge,
i.e. $[J_E,H]=0$. Based on a generalized Gibbs ensemble (GGE), which
includes the higher conserved charges, the temperature dependence of
the thermal conductivity can thus be calculated straightforwardly
\cite{SakaiKluemper,Zotos_tba_2017}.

The spin current operator $J_0$, on the other hand, is not
conserved. Whether the spin Drude weight is finite at finite
temperatures and, if so, how to calculate it analytically has been the
subject of a number of studies in the last 20 years
\cite{ZotosPrelovsek,Zotos,KluemperJPSJ,SirkerPereira,SirkerPereira2,Prosen,ProsenIlievski,ProsenNPB,PereiraPasquier,bertini_transport_2016,castro-alvaredo_emergent_2016,bulchandani_bethe-boltzmann_2018,doyon_drude_2017}. 
Based on a field theoretical treatment, a coexistence of a Drude weight
with a diffusive part was predicted for small finite temperatures
\cite{SirkerPereira2}. In frequency space, this corresponds to a Drude
peak which sits on top of a narrow Lorentzian. Further evidence for
this picture was recently obtained in a generalized hydrodynamics
equation where a diffusive term was considered as next leading
correction \cite{deNardisBernard}. In this paper we will not study the
diffusive part of the current and will instead be exclusively
concerned with the calculation of the ballistic part, i.e.~the
finite-temperature Drude weight.

The spin current density is defined by the discrete continuity equation
\begin{equation}
\label{cont}
\partial_t \sigma^z_l = -\mbox{i}[\sigma^z_l,H] = -(j_l-j_{l-1}) 
\end{equation}
from which one obtains
\begin{equation}
\label{current}
j_l = 4 i J\left(\sigma^+_l\sigma^-_{l+1} - \sigma^-_l\sigma^+_{l+1} \right)
\end{equation}
with $\sigma^{\pm} = (\sigma^x\pm i \sigma^y)/2$. The total spin
current operator is given by $J_0=\sum_l j_l$. The spin Drude weight
$D(\beta)$ at inverse temperature $\beta=1/T$ (we set $k_B=1$) can
then be defined in the following two equivalent ways. On the one hand,
one can consider the Kubo formula for the spin conductivity as a
function of frequency $\omega$
\begin{equation}
\label{Kubo}
\sigma(\omega)=\frac{\mbox{i}}{\omega}\left[\frac{\langle H_{\textrm{kin}}\rangle}{N}+\langle J_0\, ; J_0\rangle_{\textrm{ret}}(\omega)\right]
\end{equation}
where $H_{\textrm{kin}}$ is the kinetic energy operator, 
i.e. the transversal exchange terms of $H$, and $\langle
\; ; \;\rangle_{\textrm{ret}}$ is the retarded correlation
function. The real part of the conductivity is given by
\begin{equation}
\label{sigma_real}
\sigma'(\omega) = 2\pi D(\beta)\delta(\omega) + \sigma_{\textrm{reg}}(\beta,\omega).
\end{equation}
A finite Drude weight, $D(\beta)>0$, thus implies an infinite dc
conductivity. Another way to define the Drude weight is to consider
the current-current correlator directly in time $t$,
\begin{equation}
\label{Drude}
D(\beta)=\lim_{t\to\infty}\lim_{N\to\infty}\frac{ \beta \langle J_0(0) J_0(t)\rangle}{2N} = \lim_{N\to\infty}\frac{\beta}{2N}\sum_k\frac{|\langle J_0 Q_k\rangle|^2}{\langle Q_k^\dagger Q_k\rangle}.
\end{equation}
In the last step, we have projected onto a {\it complete} set of
commuting conserved charges which are orthogonal, $\langle Q_k^\dagger
Q_l\rangle = \langle Q_k^\dagger Q_k\rangle \delta_{kl}$ where
$\langle \dots\rangle$ denotes the thermal average at inverse temperature
$\beta$. If the set of charges is not complete then the r.h.s. provides a
lower bound for $D(\beta)$, the so-called Mazur bound
\cite{Mazur,Suzuki,ZotosPrelovsek}. According to this equivalent
definition, the Drude weight is the part of the current which does not
decay in time because it is protected by a finite overlap with some of
the conserved charges. The question of whether or not the XXZ chain
has a finite Drude weight at finite temperatures is an intriguing one
because $\langle J_0 Q_k\rangle =0$ for all the charges defined in
Eq.~\eqref{Charge}. This follows from simple symmetry considerations:
While the spin current \eqref{current} is odd under the spin-flip
symmetry $\sigma^z\to -\sigma^z$, all the charges in
Eq.~\eqref{Charge} are even. This puzzle was solved by realizing that
for the open XXZ chain, additional operators exist which are conserved
up to boundary terms and are odd under the spin-flip operation
\cite{Prosen,ProsenIlievski}. Later it was shown that fully conserved odd 
charges with finite overlap with the current operator can be
constructed for periodic boundary conditions
\cite{PereiraPasquier,ProsenNPB}. 

Using the Bethe ansatz there are three different approaches which have
been used so far to compute the Drude weight: (1) Starting from the
spectral representation of the Kubo formula \eqref{Kubo} and comparing
this with the change of the eigenenergies $\varepsilon_n$ of the Hamiltonian
\eqref{Ham} when threading a static magnetic flux $\Phi$ through an
XXZ ring one finds
\begin{equation}
\label{Kohn}
D=\frac{1}{2NZ}\sum_n \e^{-\beta \varepsilon_n}\frac{\partial^2\varepsilon_n(\Phi)}{\partial \Phi^2}\bigg|_{\Phi=0}
\end{equation}
with $Z$ the partition function. This is a generalization of the
Kohn formula \cite{Kohn} to finite temperatures \cite{CastellaZotos}. 
For zero temperature, in particular, the Drude weight can be 
obtained simply from the ground state energy of the system with an
added flux \cite{ShastrySutherland} leading to
\begin{equation}
\label{Dzero}
\lim_{\beta \to \infty} D(\beta) = D_{\beta \to \infty} = J\frac{\pi\sin\gamma}{2\gamma(\pi-\gamma)}.
\end{equation}
For finite inverse temperatures, the formula \eqref{Kohn} has been used in
Ref.~\cite{Zotos} to calculate $D(\beta)$ for anisotropies $\gamma=\pi/m$
on the basis of the thermodynamic Bethe ansatz (TBA). The high- and
low-temperature limits have then been analyzed in
Ref.~\cite{KluemperJPSJ}. (2) A completely different approach is based
on constructing a set of charges that have finite overlap with the
current operator and to evaluate the r.h.s.~of Eq.~\eqref{Drude}, see
Refs.~\cite{Prosen,ProsenIlievski,PereiraPasquier,ProsenNPB}. A major
difficulty in this approach is the evaluation of the correlators at
finite temperatures. So far, only the high-temperature limit has been
analyzed analytically \cite{ProsenIlievski} resulting in
\begin{equation}
\label{Dinf}
\lim_{\beta \to 0} \frac{4}{\beta}D=J^2 \frac{\sin^2(\pi n/m)}{\sin^2(\pi/m)}\left(1-\frac{m}{2\pi}\sin(2\pi/m)\right).
\end{equation}
Here the equal sign is only correct if the set of conserved charges
used is complete which is a point which is difficult to prove. For
anisotropies $\gamma=\pi/m$ it has been shown that the above result
agrees with the high-temperature limit of the TBA result obtained
using the Kohn formula. In the following we will prove that this is
also true for general anisotropies $\gamma=\pi n/m$. Note that the
Drude weight in the high-temperature limit has a fractal character
according to Eq.~\eqref{Dinf} while $D_{\beta \to \infty}$ depends smoothly on
anisotropy, see Eq.~\eqref{Dzero}. Finally (3), a third approach has
recently been proposed based on a generalized hydrodynamics (GHD)
formulation where it is conjectured that the continuity equation
\begin{equation}
\label{Cont}
\partial_t \langle q_\ell\rangle +\partial_x \langle j_\ell\rangle =0
\end{equation}
takes the form of an Euler equation\footnote{This equation is often referred to as
the Bethe-Boltzmann equation.} for the $\ell$-th quasi-particle density $\rho_\ell$
\cite{bertini_transport_2016,castro-alvaredo_emergent_2016,bulchandani_bethe-boltzmann_2018,doyon_drude_2017}
\begin{equation}
\label{BB}
\partial_t\rho_{\ell}(\theta)+\partial_x\left(v_{\ell}(\theta)\rho_{\ell}(\theta)\right)=0
\end{equation}
with effective velocity $v_{\ell}(\theta)$, where we have suppressed the time and 
space dependence. It should also be noted that both the 
particle density and effective velocity depend on position and time. The expectation
value of an extensive charge $\langle Q_m\rangle$ in a local stationary state
described by the distribution $\rho_\ell$ is given by
\begin{equation}
\label{StatState}
\langle \rho |Q_m|\rho\rangle/N = \sum_{\ell}\int d\theta\, q^m_{\ell}(\theta)\rho_{\ell}(\theta)
\end{equation}
with the subscript referring to an $\ell$-string in the BA solution
and with the superscript denoting the $m$--th bare charge
eigenvalue. If one assumes that a system which is not in equilibrium
is composed of cells which are locally described by the distribution
$\rho_{\ell}(\theta)$ then Eq.~\eqref{BB} allows to compute the time
evolution of the system along every ray $\xi=x/t$. For the
Lieb-Liniger model in the linear response regime, in particular, this
formalism has been used to obtain formulas for the expectation values
of $\langle J_n\rangle$, $\langle J_n Q_m\rangle$, and $\langle Q_n
Q_m\rangle$ \cite{doyon_drude_2017}. Formally, these results can be
straightforwardly generalized to the XXZ chain by summing over all
possible string types. An obvious question then is if the TBA formulas
for the current and current-charge expectation values obtained in this
way are exact.

To answer this question we will present in this paper a fourth
approach where we derive current and current-charge correlators
exactly starting from a generalized free energy and the operator
expression for the spin current, without using the GHD
conjecture. Based on Eq.~\eqref{Drude} we will then use these
correlators to derive a formula for the Drude weight and show that it
is identical to the GHD result and to the TBA result obtained from the
Kohn formula. Our paper is organized as follows: In Sec.~\ref{J0} we
derive exact results for $\langle J_0\rangle$ and $\langle J_0\,
Q_n\rangle$. In Sec.~\ref{DT} we obtain the Drude weight and analyze
analytically the high- and low-temperature limits for anisotropies
$\gamma=\pi n/m$. A numerical evaluation of the Drude weight for these
anisotropies and arbitrary temperatures is presented in
Sec.~\ref{Numerics}. A brief summary and conclusions are given in
Sec.~\ref{Concl}.

\section{The spin current, current-charge and charge-charge correlators}
\label{J0}
The basic object we want to consider is the reduced $n$-site density
matrix $D(n)$ obtained from the full thermal density matrix
$\rho=\exp(-\beta H )/Z$ by taking a partial trace over the other $N-n$
sites, $D(n) = \tr_{1,\cdots,N-n}\; \rho$. Note that the Hamiltonian
\eqref{Ham} is translationally invariant. The reduced density matrix
is thus only a function of the length of the segment. The elements of
the $n$-site reduced density matrix can always be expressed through a
combination of $n$-site spin correlators. For the $1$-site reduced DM,
for example, we have $1=D^1_1+D^2_2$, $\langle\sigma^z\rangle =
\tr(D(1)\sigma^z)=D^1_1-D^2_2$, $\langle \sigma^+ \rangle= D^2_1$, and $\langle \sigma^-\rangle =
D^1_2$ which allows to rewrite the matrix elements $D^\beta_\alpha$ in
terms of the expectation values of $\sigma^{z,+,-}$. Similarly, for
the $2$-site reduced DM we find $D_{12}^{21} = \langle \sigma^+_l
\sigma^-_{l+1}\rangle$. Knowing the elements of the $2$-site density 
matrix thus allows to determine the expectation value of the spin
current operator defined in Eq.~(\ref{current}). Using the Yang-Baxter
algebra, the following relation for an inhomogeneous generalization of
the reduced density matrix has already been obtained previously \cite{BoosGoehmann}
\begin{equation}
\label{DM}
D_{12}^{21}(2;\xi_1,\xi_2)-D_{21}^{12}(2;\xi_2,\xi_1)=\frac{D_1^1(1;\xi_1)- D_1^1(1;\xi_2)}{i(\xi_1-\xi_2)}.
\end{equation}
Here $\xi_i$ are spectral parameters which are put on the vertical
lines of the corresponding vertex model. Identifying the matrix
elements by the spin correlators as above we find from
\eqref{DM} the relation
\begin{equation}
\label{DM2}
\langle j_l\rangle =2 \partial_{\xi}\langle \sigma^z_l\rangle\big|_{\xi=0} \quad ; \quad \langle J_0\rangle/N =\langle j_l\rangle = -\partial_\xi \partial_{\beta h} f_\xi(\{\beta\})\big|_{\xi,h=0} .
\end{equation}
Here $f_\xi(\{\beta\})$ is the generalized free energy density with spectral
parameter $\xi$ and generalized inverse temperatures $\{\beta\}=\{\beta_0,\beta_1, \dots\}$.
It is related to the leading eigenvalue $ \Lambda(\xi) $ of the quantum transfer
matrix by 
\begin{align}
f_\xi(\{\beta\}) = - \ln \Lambda(\xi).
\end{align}
We discuss here only the transport properties of the XXZ chain at zero
magnetic field. In TBA we can write this free energy density as
\begin{equation}
\label{FreeE}
f_\xi(\{\beta\}) =-\frac{1}{2\pi}\sum_\ell\int d\theta\, p'_{\ell}(\xi-\theta)\sigma_\ell\ln[1+\eta_\ell^{-1}(\theta)] .
\end{equation}
Here $p_\ell(\theta)$ is the momentum distribution and the variables
$\sigma_\ell = \text{sign}(g_\ell)$ are the signs of auxiliary
rational numbers associated to string solutions as defined
in~\cite{takahashi_thermodynamics_1999}. For the simplest case of
anisotropy $\gamma=\pi/m$ the $g_\ell$ have a particularly simple
relation to string length $n_\ell$
\begin{equation}
\label{sgnFct}
g_\ell = m - n_\ell ,\, n_\ell = \ell \, \text{ for } \, \ell= 1, \dots, m-1 \text{ and } \, g_m =-1,\, n_m=1.
\end{equation}
The function $\eta_\ell=\rho^h_\ell/\rho_\ell$ is defined by the ratio
of hole density $\rho^h_\ell$ and particle density $\rho_\ell$ of the
$\ell$-th particle (string). It fulfills the TBA equations
\begin{align}
\ln\eta_\ell(\theta) &=\sum_n \beta_n q^n_\ell(\theta) + \sum_\kappa \int d\mu K_{\ell \kappa}(\theta-\mu) \sigma_\kappa  \ln(1+\eta_\kappa^{-1}(\mu)),\nonumber\\
\label{basic}
&\equiv\sum_n \beta_n q^n_\ell + \left[K * \sigma \ln(1+\eta^{-1}) \right]_\ell
\end{align}
with charges $q^n_\ell$, Lagrange multipliers (generalized
temperatures) $\beta_n$, an integration kernel $K$, and '$*$' denoting
a convolution and sum over Bethe strings. For the first few charges we
have, in particular, $\beta_0=\beta h$, $q^0_\ell = n_\ell$, and
$\beta_1=\beta$, $\gamma(4J\sin\gamma)^{-1}\varepsilon_\ell =
\partial_\theta p_\ell= p'_\ell$. In the following, we rescale the
energy $\gamma(4J\sin\gamma)^{-1}\varepsilon_\ell\to
\varepsilon_\ell=q^1_\ell$ to absorb the scaling factor. Furthermore,
we use the shorthand notation $\partial_n\equiv
\partial_{\beta_n}$. Dressed charges $\dr{q}^n_\ell$ 
are defined by the integral relation
\begin{equation}
\label{dressIntFn}
\dr{q}^n_\ell = q^n_\ell - \left[ K * \sigma \vartheta \dr{q}^n\right]_\ell  \, 
\end{equation}
where we have defined the Fermi factor
$\vartheta_\ell=1/(1+\eta_\ell)=\rho_\ell/(\rho_\ell +
\rho^h_\ell)$.  It is also very useful to realize the following simple
relation of the dressed charges to logarithmic derivatives
of the $\eta$-functions
\begin{equation}
\label{eta}
\partial_{n} \log \eta_\ell(\theta) = \dr{q}_\ell^n(\theta)\, .
\end{equation}
In order to calculate the expectation value of the current $\langle J_0\rangle$ we note that
\begin{equation}
\label{eta2}
\partial_{0}\ln(1+\eta_\ell^{-1})=-\frac{\partial_{0}\ln\eta_\ell}{1+\eta_\ell}=-\vartheta_\ell \dr{q}_\ell^0  
\end{equation} 
leading to
\begin{equation}
\label{J0e1}
\langle J_0\rangle /N = -\partial_\xi \partial_{0} f_\xi(\{\beta\})\big|_{\xi=0} = \frac{1}{2\pi }\sum_\ell \int d\theta \,\sigma_\ell \varepsilon'_\ell(\theta) \vartheta_\ell(\theta) \tilde{q}_\ell^0(\theta)\, ,
\end{equation}
where we have used that $p''(\theta) = \varepsilon'(\theta)$. There
are various ways to rewrite this equation. Here we want to bring it
into a form similar to the one conjectured within the GHD
approach. The basic identity we want to make use of is
\begin{align}
\begin{split}\label{identity}
 {}&\sum_{\ell}\int d\theta \left[K*\sigma \partial_n\ln(1+\eta^{-1})\right]_\ell \sigma_\ell \partial_m \ln(1+\eta_\ell^{-1})\\
&=\sum_\ell \int d\theta \left[K* \sigma \partial_m\ln(1+\eta^{-1})\right]_\ell \sigma_\ell \partial_n \ln(1+\eta_\ell^{-1}).
\end{split}\end{align}
Using Eq.~\eqref{basic} we can express
$K*\sigma_\ell \partial_0\ln(1+\eta_\ell^{-1})= \partial_0  \ln\eta_\ell -q_\ell^0$
and $K*\sigma_\ell \partial_2\ln(1+\eta_\ell^{-1})=\partial_2\ln\eta_\ell
-q_\ell^2$. In this case the identity \eqref{identity} yields
\begin{equation}
\label{identity2}
\sum_\ell \int d\theta\,q^0_\ell \sigma_\ell  \underbrace{\partial_2\ln(1+\eta_\ell^{-1})}_{-\vartheta_\ell\dr{q}_\ell^2} =\sum_\ell  \int d\theta\, q_l^2 \sigma_\ell \underbrace{\partial_0\ln(1+\eta_\ell^{-1})}_{-\vartheta_\ell \dr{q}_\ell^0}
\end{equation}
The expectation value of the current operator \eqref{J0e1} can thus also be written as
\begin{equation}
\label{J0e2}
\langle J_0\rangle/N = \frac{1}{2\pi}\sum_\ell \int d\theta\, \sigma_\ell\vartheta_\ell \underbrace{\dr{q}_\ell^2}_{=\dr{\partial_\theta \varepsilon_\ell}} \underbrace{q_\ell^0}_{=n_\ell} = \sum_\ell \int d\theta \, v_\ell(\theta) \rho_\ell(\theta) {q}_\ell^0(\theta) 
\end{equation}
where the rapidity density $\dr{\partial_\theta p}_\ell = 2\pi\sigma_\ell (\rho_\ell +
\rho^h_\ell)$ and effective velocity $ v_\ell \equiv
{\dr{\partial_\theta \varepsilon}_\ell}/{\dr{\partial_\theta  p}_\ell}$  are defined by the dressed
derivatives with respect to the spectral parameter of the energy and
the momentum. This formula agrees with the conjectured general 
current formula used in GHD
and appearing in Ref.~\cite{bertini_transport_2016,castro-alvaredo_emergent_2016}.

The correlator $\langle J_0 Q_n\rangle/N$ can also be computed from the
free energy $f_\xi(\{\beta\})$ defined in Eq.~\eqref{FreeE} via derivatives
with respect to the appropriate Lagrange multiplier $\beta_j$, see
Eq.~\eqref{eta}. We find in particular,
\begin{equation}
\label{JQ}
\langle J_0 Q_n\rangle/N = -\partial_\xi\partial_{0}\partial_{n} f_\xi(\{\beta\}) = -\frac{1}{2\pi}\sum_\ell\int d\theta\,(\partial_\theta \varepsilon_\ell)\sigma_\ell \partial_{0}\partial_{n}\ln(1+\eta_\ell^{-1}). 
\end{equation}
In order to simplify this result we use the following relation
\begin{equation}
\label{JQ_relation}
\sum_\ell \int d\theta\, q_\ell^k \sigma_\ell \partial_m\partial_n\ln(1+\eta_\ell^{-1})= \sum_\ell \int d\theta\, \sigma_\ell \vartheta_\ell(1-\vartheta_\ell)\dr{q}^k_\ell\dr{q}^m_\ell\dr{q}^n_\ell\, ,
\end{equation}
which is proven in Appendix \ref{App_JQ}. Using this relation for the charge-current correlator \eqref{JQ} leads to our final result
\begin{equation}
\label{JQfinal}
\langle J_0 Q_n\rangle/N = -\sum_\ell \int \frac{d\theta}{2\pi}\dr{\partial_\theta \varepsilon}_\ell\sigma_\ell \vartheta_\ell (1-\vartheta_\ell) \dr{q}^0_l \dr{q}^n_\ell =-\sum_\ell \int d\theta v_\ell \rho_\ell (1-\vartheta_\ell)\dr{q}_\ell^0\dr{q}_\ell^n \, ,
\end{equation}
where we have once more made use of the rapidity density and effective velocity relations.
As above, this result is consistent with a generalization of the formula in
Ref.~\cite{doyon_drude_2017} from the Lieb-Liniger model to the case of multiple particle species.
Analogously, the charge-charge correlator is
given by
\begin{eqnarray}
\label{QQ}
\langle Q_n Q_m\rangle /N &=& -\partial_{n}\partial_{m} f_{\xi=0}(\{\beta_n\}) =\frac{1}{2\pi}\sum_\ell\int d\theta\, (\partial_\theta p_\ell)\sigma_\ell\partial_{n}\partial_{m}\ln(1+\eta_\ell^{-1}), \nonumber \\
&=& \frac{1}{2\pi}\sum_\ell\int d\theta\,\sigma_\ell\vartheta_\ell(1-\vartheta_\ell)\dr{\partial_\theta p}_\ell\dr{q}_\ell^n\dr{q}_\ell^m, \nonumber \\
&=& \sum_\ell \int d\theta\, \rho_\ell (1-\vartheta_\ell) \dr{q}^n_\ell \dr{q}^m_\ell \, ,
\end{eqnarray}
where we have used again the relation \eqref{JQ_relation} in the
second step. For the special case $Q_n=Q_m=Q_2=J_E$ this reproduces
the formula needed to calculate the thermal Drude weight by TBA first
derived in
\cite{Zotos_tba_2017}. If we only take a single derivative, then we obtain a TBA formula for the energy current
\begin{equation}
\label{Ecurrent}
\langle J_E\rangle = -\partial_2 f_{\xi=0}(\{\beta\})=\frac{1}{2\pi}\sum_\ell\int d\theta\, p'_\ell \sigma_\ell \vartheta_\ell \tilde{q}_\ell^2=\sum_\ell\int d\theta\, \rho_\ell q_\ell^2 = \sum_\ell\int d\theta\, v_\ell \rho_\ell q_\ell^1 \, .
\end{equation}
This result agrees with Eq.~\eqref{StatState} and also with
Eq.~\eqref{J0e2} provided we replace $q^0_\ell$ with $q^1_\ell$. 

\section{The Drude weight}
\label{DT}
Using the expressions for the spin-charge and charge-charge
correlators in Eq.~\eqref{JQfinal} and Eq.~\eqref{QQ} of the previous
section, the Drude weight \eqref{Drude} can be determined from
\begin{equation}
\label{Drude2}
D= \lim_{N\to\infty}\frac{\beta}{2N}\sum_n\frac{\langle J_0 Q_n\rangle^2}{\langle Q_n^2\rangle} =\frac{\beta}{2}\sum_n\frac{\left(\sum_\ell \int d\theta v_\ell \rho_\ell (1-\vartheta_\ell)\dr{q}_\ell^0\dr{q}_\ell^n\right)^2}{\sum_\ell \int d\theta \rho_\ell (1-\vartheta_\ell) \dr{q}^n_l \dr{q}^n_\ell} \, .
\end{equation}
In order to simplify Eq.~\eqref{Drude2} 
we follow the original argument by Mazur \cite{Mazur} and define a quantity 
$Z=J_0-\sum_n c_n Q_n$ with $\langle Z^2\rangle \geq 0$, and the set $Q_n$ being the 
complete set of conserved charges. This leads to the relation\footnote{Taking $\langle J_0 J_0 \rangle$ as shorthand for $\lim_{t \to \infty} \langle J_0(0) J_0(t) \rangle$.}
\begin{align}
\label{eq:mazurBoundProof}
\langle J_0  J_0 \rangle \geq 2 \sum_{n} c_n \langle J_0  Q_n \rangle- \sum_{n,m} c_n c_m \langle  Q_n   Q_m \rangle \, .
\end{align}
Maximizing the right hand side with respect to the vector $\vec{c}$ leads to the condition
\begin{equation}
\label{eq:constCondition}
\sum_{n} c_n \langle Q_n  Q_m \rangle = \langle J_0  Q_m \rangle.
\end{equation}
Using the expressions \eqref{JQfinal} and \eqref{QQ} after bringing the overall constants to the LHS
\begin{eqnarray}
\label{completeSetCondition}
&& \sum_{n,\ell} c_n \int d\theta \rho_\ell(\theta) (1-\vartheta_\ell(\theta)) \dr{q}^n_\ell(\theta) \dr{q}_\ell^m(\theta) =-\sum_\ell\int d\theta v_\ell(\theta) \rho_\ell(\theta) (1-\vartheta_\ell(\theta)) \dr{q}_\ell^0 \dr{q}_\ell^m(\theta) \nonumber\\
\Leftrightarrow && \sum_\ell\int d\theta \rho_\ell(\theta) (1-\vartheta_\ell(\theta)) \left( \sum_n c_n \dr{q}_\ell^n(\theta)  + v_\ell(\theta) \dr{q}_\ell^0 \right) \dr{q}_\ell^m(\theta) = 0 \, . 
\end{eqnarray}
Next, we use that $\{q^n\}$ is a complete set of conserved charges
with non-vanishing overlap with the spin current. This set comprises
of the quasi-local charges~\cite{ProsenIlievski} and additional
charges $Q_1,Q_2,\cdots$. The exact form of these additional charges
does not matter as long as they make the set complete. If this is the
case, then the Mazur argument can be applied. We will see in the
following that the additional charges drop out in the final result. We
further also assume completeness in the sense that the vanishing of
the sum-integral of Eq.~\eqref{completeSetCondition} for any charge
automatically implies the vanishing of the integrand,
\begin{align}
\label{eq:constEquality}
\sum_n c_n \dr{q}_\ell^n(\theta)   = -v_\ell(\theta) \dr{q}_\ell^0 \, .
\end{align}
Under these assumptions, the bound obtained should be exhaustive and
we find the following expression for the conserved part of the spin
current
\begin{align}
\label{Drudefinal}
2D \beta^{-1}=\langle J_0 J_0\rangle/N &= \sum_{n} c_n\langle J_0  Q_n \rangle/N - \sum_{n,m} c_n c_m \langle  Q_n   Q_m \rangle/N =\sum_{n} c_n \langle J_0  Q_n \rangle/N \, , \nonumber\\
 &= - \sum_\ell\int d\theta\, \sum_n c_n \rho_\ell(\theta) (1-\vartheta_\ell(\theta)) \dr{q}_\ell^n(\theta)\, v_\ell(\theta) \dr{q}_\ell^0(\theta)  ,\nonumber\\
&=\sum_\ell\int d\theta \rho_\ell(\theta) (1-\vartheta_\ell(\theta)) \left[v_\ell(\theta) \dr{q}_\ell^0(\theta)\right]^2.
\end{align}
The last part of our derivation is based on the same assumptions used
in~\cite{doyon_drude_2017}. Importantly however, the expressions for
current and current-charge expectation values are derived from first
principles.

\subsection{Equivalence with the Drude weight formula by Zotos}
\label{Zotos}
Starting from \eqref{Drudefinal} it is now straightforward to show
that our result is identical to the one obtained 20 years ago based on
the Kohn formula and using the TBA to calculate the curvature of
energy levels \cite{Zotos,KluemperJPSJ}. Rewriting the
particle density and filling fraction in terms of $\eta$-functions we
obtain
\begin{eqnarray}
\label{Zotos1}
D &=&\frac{\beta}{2}\sum_\ell\int d\theta\, \frac{\rho_\ell+\rho_\ell^h}{(1+\eta_\ell)(1+\eta_\ell^{-1})}\left(\dr{\partial_\theta\varepsilon}/\dr{\partial_\theta p}\right)^2 (\dr{q}_\ell^0)^2 \nonumber \\
&=& \frac{\beta}{4\pi}\sum_\ell\int d\theta\, \sigma_\ell \frac{(\dr{q}_\ell^2)^2(\dr{q}_\ell^0)^2}{\dr{q}_\ell^1(1+\eta_\ell)(1+\eta_\ell^{-1})}
\end{eqnarray}
Now we can use the relation $\partial_\theta\ln\eta_\ell = \beta \dr
{\partial_\theta \varepsilon}_\ell= \beta\dr{q}_\ell^2$ to obtain---up
to a normalization factor--- the well-known result
\cite{KluemperJPSJ}
\begin{equation}
\label{Zotos2}
D=\frac{1}{4\pi \beta }\sum_\ell\int d\theta\,\sigma_\ell\frac{(\partial_\theta\ln\eta_\ell)^2(\partial_0\ln\eta_\ell)^2}{(\partial_1\ln\eta_\ell)(1+\eta_\ell)(1+\eta_\ell^{-1})}\, .
\end{equation}
Note that by restoring the scaling factor $4J\sin(\gamma)/\gamma$ the
result in \eqref{Zotos2} agrees with \cite{KluemperJPSJ}.

Here we have thus provided an alternative derivation of the Drude
formula which makes use of an (exhaustive) Mazur bound and first
principles derivations of current correlators instead of the Kohn
formula. Note, however, that both approaches use the TBA formalism so
the rederivation presented here should not be understood as being
completely independent.

\subsection{Low-temperature limit}
\label{LowT}
The low-temperature asymptotics of Eq.~\eqref{Zotos2} have already
been determined in~\cite{ShastrySutherland,zvyaginLowT,zvyaginHubbard} with
\begin{align}
\label{lowTeqn}
D_{\beta\rightarrow \infty} = J \frac{\pi \sin\gamma}{2 \gamma (\pi - \gamma)} 
\end{align}
consistent with the known zero temperature result \eqref{Dzero}. Our
numerical data discussed in more detail in Sec.~\ref{Numerics} also
agree with this low-temperature formula, up to the point where the
numerics breaks down, see Fig.~\ref{Fig3}.

We note, furthermore, that this formula also follows directly from the
alternative expression \eqref{Drudefinal} by observing that the
particle/ hole densities vanish around the origin. So only regions
with constant effective velocity $v_\pm = \pm
2 J\pi\sin(\gamma)/\gamma$ have non-zero particle/ hole density. Then
taking into account that \eqref{JQfinal} reduces to $(v_\pm)^2$
multiplied by $1/2$ times the zero field susceptibility, $\chi_0 = 
\frac{1}{4J \pi(\pi-\gamma)}\frac{\gamma}{\sin\gamma}$, Eq.~\eqref{lowTeqn} 
follows provided that one also reintroduces the rescaling factor
$4J\sin(\gamma)/\gamma$.

\subsection{High-temperature limit}
\label{HighT}
The high temperature asymptotics of the Drude weight \eqref{Dinf} has
been obtained by constructing families of quasi-local charges
\cite{ProsenIlievski}. Numerics based on the GHD approach
agree with this bound~\cite{ilievski_microscopic_2017} and analytical
GHD calculations for certain density and current profiles reproduce
it~\cite{collura_analytic_2018}. A proof for~\eqref{Dinf} directly
from the quantum transfer matrix approach is known only for $\gamma
=\frac{
\pi}{m}$ \cite{KluemperJPSJ}. We generalize this transfer matrix result 
to anisotropies $\gamma = \frac{n\pi}{m}$ making use of the Y-system decomposition
in~\cite{kuniba_continued_1998}, and the usual unscaled temperatures appearing therein.
A rational $\pi/\gamma$ can be written as a continued fraction of length $\alpha$ determined by integers $\nu_j$.
There are $L = \sum^\alpha_{j=1} \nu_j$ functional equations for $\eta_\ell$ 
terms. Importantly the final two `boundary' $\eta$ are given by
\begin{align}
\label{yDecomp}
\eta_{L -1}(x) = e^{\beta h m /2}K(x)\, , && \eta_{L }(x) = e^{\beta h m /2}\frac{1}{K(x)}\, .
\end{align}
These boundary $\eta$ are the only terms with 
magnetization appearing in odd powers, meaning that 
\begin{align}
\partial_{\beta h} \eta_{L -1} \big|_{h=0} = \partial_{\beta h} \eta_{L} \big|_{h=0} = \frac{m}{2}\, ,&&  \text{and} && \partial_{\beta h} \eta_j \big|_{h=0} = 0, && \text{for}&& 1 \leq j \leq L -2\, .
\end{align}
Thus only the final two $\eta$ terms contribute to the Drude
weight at vanishing field. We denote the boundary string pair as a particle
($\eta_{L-1}$)/ hole ($\eta_{L})$ pair with string lengths
$\mu := n_{L-1}$ and $\bar{\mu} := n_{L}$ respectively. From Eq.~\eqref{yDecomp}, this  
pair is determined by $K(x)$, with these $K(x)$ expressible in terms of transfer matrices $T_{r -1}(x)$,
\begin{align}
\label{kDefinition}
K(x) =  \frac{T_{\mu -1} (x + i p_0 w_0) }{ T_{\bar{\mu} -1} (x + i(m + p_0 w_0)) }\, ,
\end{align}
with $w_0 p_0 = (-1)^{\alpha+1} p_L +p_0-2 \bar{\mu}$, where $\alpha$ is
the length of our continued fraction, $p_0 = \pi/\gamma$, and $p_L =
\pi/(\gamma m)$. By abuse of notation we express the eigenvalues of
$T_{r-1}$ as
\begin{align}
Q(x) &= \prod_{j=1}^M \sinh\left( \frac{\gamma}{2}(x - \omega_j) \right)\, ,\\
\phi^\pm(x) &= \left\{ \sinh\left(\frac{\gamma}{2} (x\pm i u)\right) \right\}^{\frac{N}{2}}\, , \text{ where } u = - \frac{4 J \beta  \sin(\gamma)}{\gamma N}\, , \\
\label{tEigenvalues}
T_{r -1}(x) &= Q(x + i r) Q(x - i r) \sum_{j =1}^r \frac{\phi^-[x+i(2j-2 -r)]\phi^+[x+i(2j -r)]}{Q[x + i(2j -2 -r)]Q[x+i(2j-r)]}.
\end{align}
By use of both the functional relation, which is valid
for rational values of $\pi/\gamma$,
\begin{align} 
T_{\mu + 2\bar{\mu} -1}(x) = T_{\mu -1}(x) + 2 T_{\bar{\mu}-1}(x +i (\mu+\bar{\mu}))\, ,
\end{align}
and periodicity conditions of $Q(x)$ and $\phi(x)$ the
sums of Eq.~\eqref{tEigenvalues} inserted into Eq.~\eqref{kDefinition} 
simplify to
\begin{align}
\label{simpleKvalue}
K(x)+1 = m \left(\sum_{j=1}^{\bar{n}} \frac{\phi^-[x+i(p_0 w_0+2j-2 )]\phi^+[x+i(p_0 w_0+ 2j)]}{Q[x + i(p_0 w_0+2j -2 )]Q[x+i(p_0 w_0+2j)]}\right)^{-1}\, .
\end{align}

In this form the Trotter limit at infinite temperature $\beta \rightarrow 0$ can 
be used to determine the first order temperature effect, by noting that in 
this limit the Bethe roots can be identified identically with zero. 
For brevity take $\xi_j(x) = \coth\left(\frac{\gamma x}{2} + i \gamma j\right)$ and the first order
in $\beta$ yields (with $x' = x+ i w_0 p_0$)
\begin{align}
\frac{\phi^-[x'+i(2j-2 )]\phi^+[x'+i( 2j)]}{Q[x' + i(2j -2 )]Q[x'+i(2j)]} = \left( 1 + i J \sin{\gamma}\beta \left(\xi_{j-1}(x')-\xi_j(x')\right)\right)\, .
\end{align}
From this expansion it is straightforward to complete
the sum in the denominator of Eq.~\eqref{simpleKvalue}. Expanding
again in $\beta$ leads to the first order correction
\begin{align}
K(x) +1 = \frac{m}{\bar{\mu}} \left( 1 -  \frac{i J \sin{\gamma}}{ \bar{\mu}} \beta \left(\xi_{0}(x+ iw_0 p_0)-\xi_{\bar{\mu}}(x+i w_0 p_0)\right)\right) + O(\beta^2)\, .
\end{align}

This result can then be inserted into the Drude weight formula 
Eq.~\eqref{Zotos2}, which reduces to the integral
\begin{align}
D_{\beta \to 0}= -\frac{i \beta J^2 \sin^2(\gamma)}{8 \pi \alpha} m \int d\theta \left(\frac{(\partial_\theta \xi_0(2\theta/\gamma + i w_0 p_0) - \partial_\theta \xi_{\bar{\mu}}(2\theta/\gamma +i w_0 p_0) )^2}{\xi_0(2\theta/\gamma + i w_0 p_0) - \xi_{\bar{\mu}}(2\theta/\gamma +i w_0 p_0) }\right)\, .
\end{align}
This can then be integrated to obtain the leading order corrections of the  high temperature result
\begin{align}
\label{highT_Prosen}
4\beta^{-1}D_{\beta \to 0} = J^2 \frac{ \sin^2(\gamma)}{\sin^2( \gamma p_\alpha)} \left( \frac{ \gamma p_\alpha m}{\pi} - \frac{m}{2 \pi} \sin( 2 \gamma p_\alpha) \right) + O(\beta^2),
\end{align}
where the $O(\beta)$ term is found to vanish. This is exactly the
Prosen bound \eqref{Dinf} as found via the construction of quasi-local
charges in~\cite{PereiraPasquier,ProsenNPB} provided $p_\alpha =
\pi/(\gamma m)$, which is proven in Appendix
\ref{Frac_Id} by induction.

\section{Numerical evaluation of $D(\beta)$ for arbitrary temperatures}
\label{Numerics}
In order to obtain the Drude weight, two numerical schemes were
used. The first was used as a check and involves the preparation of
two spin chains at thermal equilibrium with some small magnetic field
difference between the two, which are then joined at the
origin. The system is evolved via the Euler relations
\eqref{BB}, which permit a linear response calculation of the Drude weight. This
first method has been applied to this problem previously in
Refs.~\cite{bulchandani_bethe-boltzmann_2018,ilievski_microscopic_2017}. The
second method involves the explicit evaluation of~\eqref{Drudefinal},
which can be computed much more quickly and was analytically shown
in~\cite{doyon_drude_2017} to be equivalent to the first method.

Both methods involved determining the Fermi-weights
$\vartheta_\ell(\theta) = \frac{1} {1 + \eta_\ell(\theta)}$ via the
Yang-Yang method by obtaining the hole/ particle density ratio
$\eta_\ell(\theta)$ via Eq. \eqref{basic}. With an initial guess
function $M_\ell(\theta)$ the calculation was carried out by simple
half step updates until it reached the desired convergence. Explicitly
this was carried out by the following steps
\begin{itemize}
\item Use $\eta^{N}_\ell(\theta)$ to compute the set of 
$M_\ell^{N}(\theta)= \log \left( 1 + \frac{1}{\eta_\ell^{N}(\theta)} \right)$.
\item Take the Fast Fourier Transform (FFT) of $M_\ell^N(\theta)$.
\item Solve the transformed form of Eq.~(\ref{basic}) for the dressed energy, $\text{FFT}(\dr{\varepsilon}^{\text{temp}}_\ell(\theta))$.
\item Invert the FFT and update $\eta_\ell^{N+1} = (\eta_\ell^{\text{temp}} + \eta_\ell^{N})/2$.
\item Return to the first step with the updated guess $M_{\ell}^{N+1}$.
\end{itemize}

Once the functions $\eta_\ell(\theta)$ have converged, the dressed
charges can be obtained using the relation \eqref{eta}. The dressed
spin is known in the zero field limit to be $\dr{q}^0_\ell = 0$ for
$\ell = 1 \dots L-2$ and $\dr{q}^0_{L-1}=\dr{q}^0_L=m/2$ with $\gamma
= \frac{n \pi}{m}$. Note that the $n$ appearing in the anisotropy is
not connected to the string length $n_\ell=q^0_\ell$.  This provides a
first check on the validity of the solution.

\begin{figure}[h]
\begin{center}
\includegraphics[width=0.7\columnwidth]{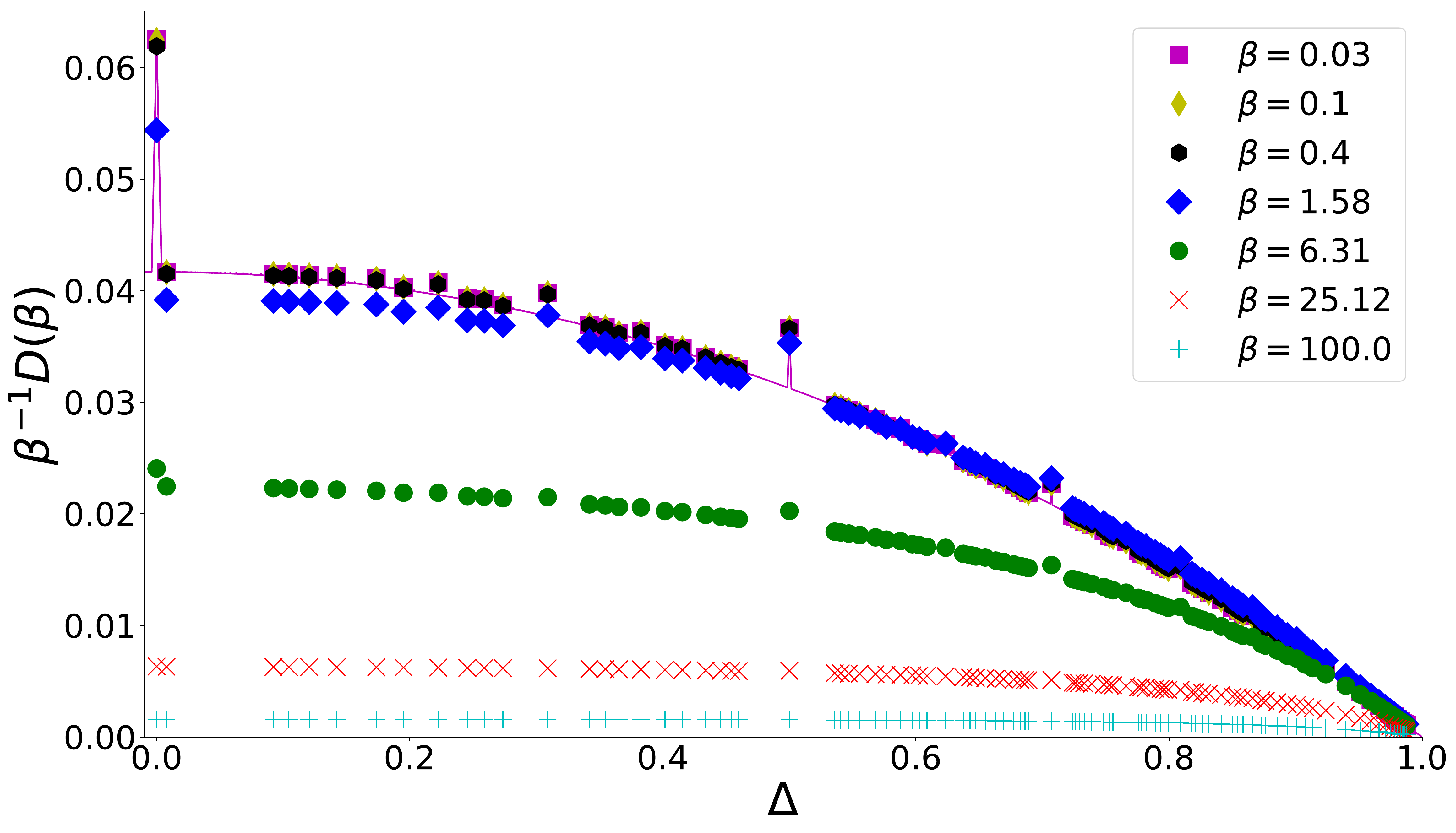}
\end{center}
\caption{Drude weight coefficient $\beta^{-1}D(\gamma,\beta)$ for various anisotropies $\gamma=\pi n/m$ and temperatures. Note that the high temperature results ($\beta=0.4,0.1,0.03$) are partly on top of each other on this scale and agree with the analytical infinite temperature result (solid line). $D(\beta)$ is a nowhere continuous function except for at $\beta^{-1}=0$. Note that the change of $\beta^{-1}D$ with decreasing temperature is not uniform: the data for $\beta=0.4$ and $\beta=1.58$ show a crossover at $\cos(\gamma) = \Delta \approx 0.59$.}
\label{Fig1}
\end{figure}

A first question we want to address numerically is how the nowhere
continuous bound for the Drude weight \eqref{Dinf} evolves into the
zero temperature Drude weight \eqref{Dzero} which is a smooth function
of anisotropy. From Fig.~\ref{Fig1} it becomes clear 
that $D(\beta)$ is in fact a fractal for any finite temperature.
Once the part of the current which is not protected by conservation
laws starts to relax due to finite-temperature Umklapp scattering, the
structure of the conserved charges odd under spin-flip
symmetry---which strongly depends on the anisotropy $\Delta=\cos(\pi
n/m)$---becomes visible in the remaining Drude weight. The $\beta ^{-1}=0$ case
is special because Umklapp scattering is an irrelevant operator. There
is no mechanism for current relaxation in a completely clean system at
zero temperature and the integrable structure of the model, which is
responsible for the discontinuous $D(\beta>0)$ as a function of
anisotropy, plays no role.

\begin{figure}
\begin{center}
 \includegraphics[width=0.48\columnwidth]{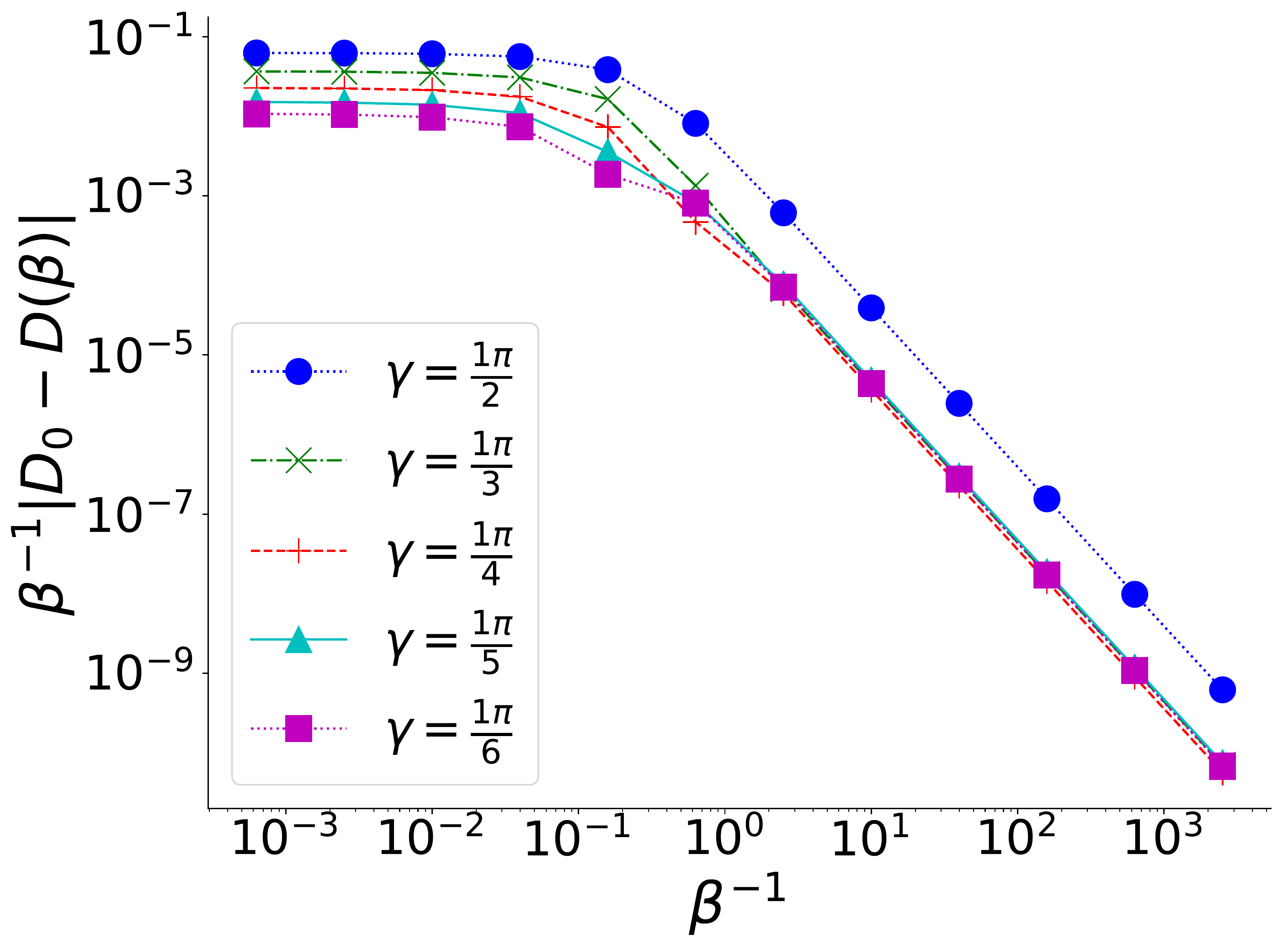}
\includegraphics[width=0.48\columnwidth]{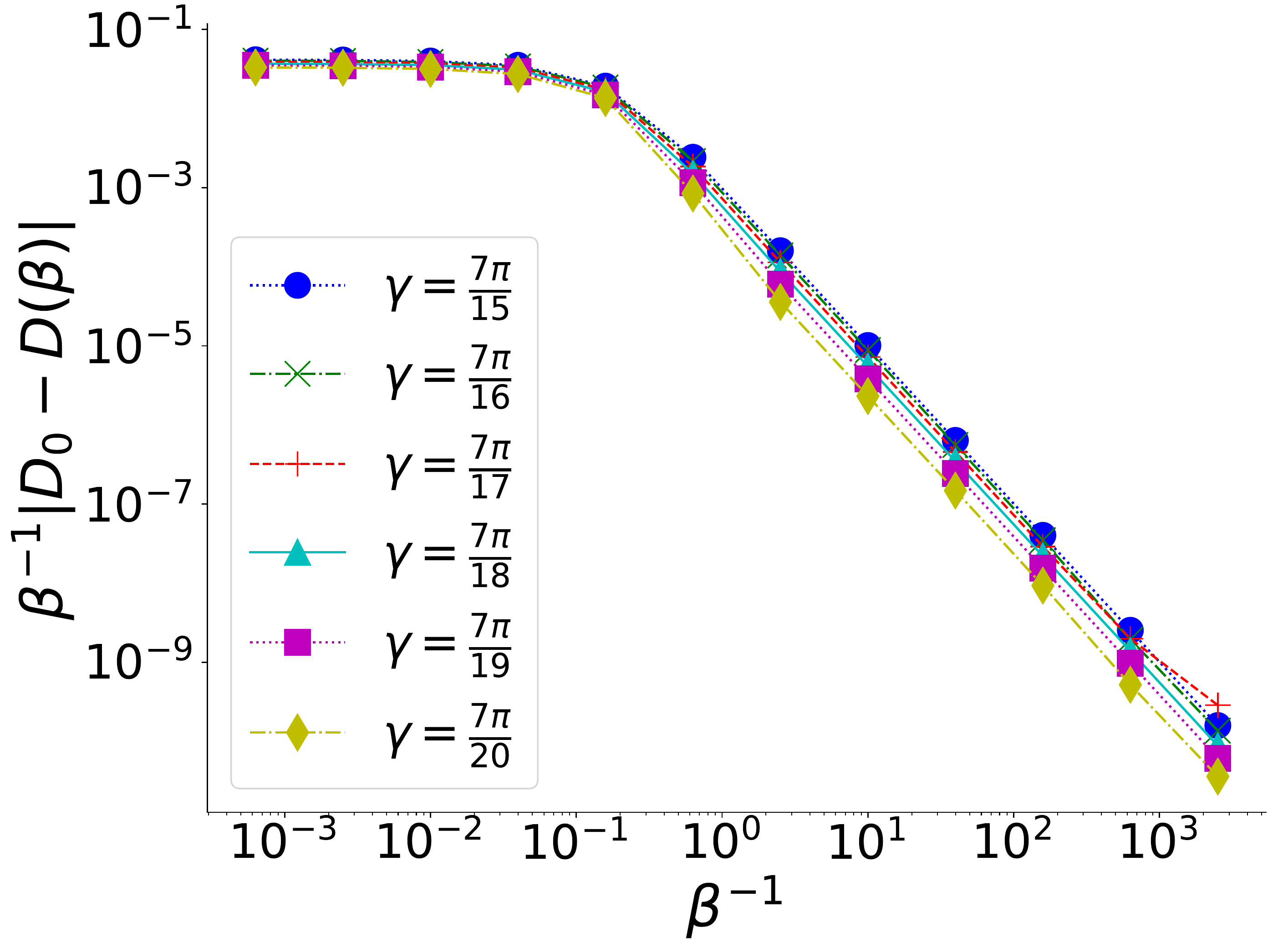}
\end{center}
\caption{\label{fig:prosenBDfig}Absolute difference between the Prosen bound $\beta^{-1} D_{\beta \to 0}$ of \eqref{Dinf} and the Drude weight coefficient $\beta^{-1}D$ demonstrating convergence to the Prosen bound. Note the crossover between curves for different anisotropies at $\beta^{-1} \approx 0.5$ in the left panel. A power-law scaling consistent with $|D(\beta) - D_{\beta \to 0}| \sim \beta^3$ at high temperatures is observed, agreeing with the TBA result see Eq.~\eqref{highT_Prosen}.}
\label{Fig2}
 \end{figure}

Next, we want to consider the high-temperature limit in more detail. In Fig.~\ref{Fig2} the difference between $\beta^{-1}D(\beta)$ and the bound \eqref{Dinf} is shown.
 We note first that for both sets of anisotropies,
 $\gamma=\pi/m$ and $\gamma=7\pi/m$, the numerical data show a
 power-law decay in temperature towards the high-temperature
 bound. Interestingly, the Drude weights at a given temperature order
 differently as a function of anisotropy in the case $\gamma=\pi/m$
 for temperatures above and below $\beta^{-1}\approx 0.5$.

\begin{figure}
\begin{center}
\includegraphics[width=0.7\columnwidth]{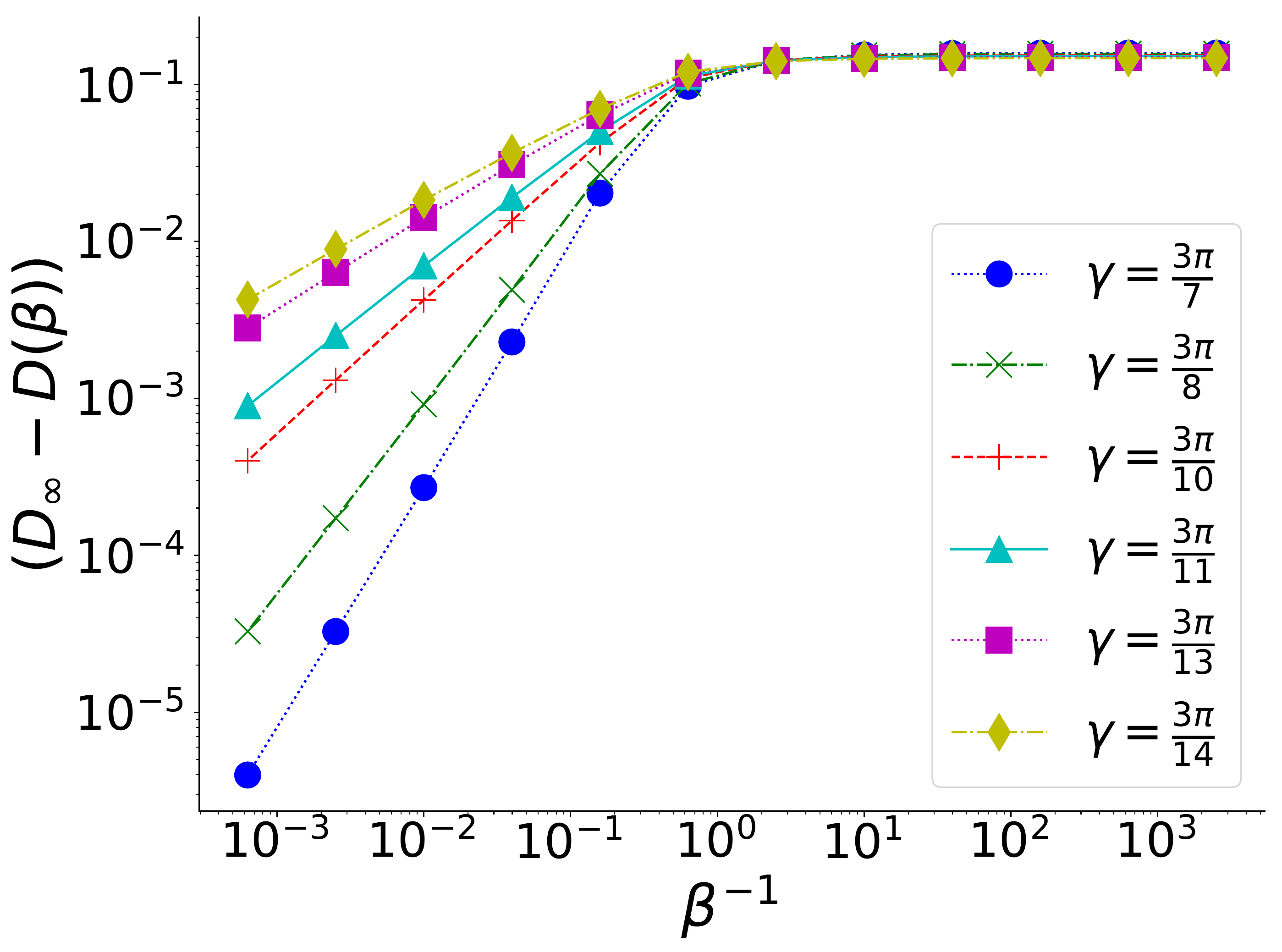}
\end{center}
\caption{The numerical data at low temperatures show a power-law scaling consistent with $D_{\beta \to \infty}- D(\beta)  \sim \beta^{2\gamma/(\gamma-\pi)}$.}
\label{Fig3} 
\end{figure}

Finally, we also want to consider the low-temperature limit for
general anisotropies $\gamma=\pi n/m$. In Ref.~\cite{Zotos}
it was observed that the Drude weight at low temperatures scales as
$D(\beta)\sim D_{\beta \to \infty}-\alpha \beta^{2/(1-m)}$ for anisotropies $\gamma=\pi/m$
with some constant $\alpha$. If this scaling does hold for all
commensurate anisotropies then one would expect $D(\beta)\sim D_{\beta \to \infty}
-\alpha \beta^{2\gamma/(\gamma-\pi)}$. In Fig.~\ref{Fig3} we show
exemplarily for anisotropies $\gamma=3\pi/m$ that this expectation is
consistent with our numerical data.
 
\section{Conclusions}
\label{Concl}
The main purpose of this paper was to provide a first principles
derivation for the expectation value of the spin current as well as
current-charge and charge-charge expectation values in steady states
described by given particle and hole distributions within the TBA
approach. The main ingredient to derive exact formulas for these
quantities was to relate the spin current with a matrix element of the
two-site reduced density matrix. We then used the fact that this
matrix element can be obtained from a generalized free energy by
taking a derivative with respect to a spectral parameter while the
charges were generated by taking derivatives with respect to the
Lagrange parameters (generalized temperatures) $\beta_n$. We showed
that the results derived in this way are consistent with
a multi-particle generalization of known Lieb-Liniger results as conjectured
in~\cite{doyon_drude_2017} and hence with the formula in~\cite{Zotos}. 
Using the Mazur bound and assuming that it becomes
exhausted if one considers the full TBA particle content we also
derived a closed-form expression for the spin Drude
weight. Straightforward manipulations showed that our result is
identical to the TBA result obtained 20 years ago based on calculating
the curvature of energy levels and using the finite-temperature Kohn
formula. While consistent results for the Drude weight have now been
obtained by the Kohn formula, by constructing the quasi-local charges
protecting the Drude weight, and by the approach presented in this
paper we would like to stress that all of these results are limited to
the commensurate anisotropies $\gamma=\pi n/m$ and make use of the TBA
formalism. While the construction of quasi-local charges has provided
a definitive finite lower bound for anisotropies $|\Delta|<1$ it is,
in our view, still not completely excluded that parts of the Drude
weight are missed in these approaches. In particular, it seems to us
that we are not at the point yet where we can claim that a fractal
structure of $D(\beta>0)$ has been fully proven.

Putting such fundamental questions aside and taking the TBA result
$D(\beta)$ as given, we have extended the analysis of the
high-temperature asymptotics from the case $\gamma=\pi/m$ to all
commensurate anisotropies $\gamma=\pi n/m$. Our analytical result in
the high-temperature limit is identical to the bound obtained
previously by considering the contribution of all known quasi-local
charges. Finally, we have also presented a numerical evaluation of
$D(\beta)$ for all temperatures showing that the TBA Drude weight has
fractal character for all finite temperatures and that the
low-temperature scaling follows a power law with exponent
$D_{\beta \to \infty}- D(\beta) \sim \beta^{2\gamma/(\gamma-\pi)}$.

\section*{Acknowledgements and funding information}
The authors acknowledge support by the Deutsche Forschungsgemeinschaft
(DFG) via Research Unit FOR 2316. JS acknowledges support by the
Natural Sciences and Engineering Research Council (NSERC, Canada).


\begin{appendix}
\section{Proof of identity (\ref{JQ_relation})}
\label{App_JQ}
The identity \eqref{JQ_relation}, which we want to prove here, can
also be written as
\begin{equation}
\label{App1e1}
\int d\theta\, \sigma  q^k \partial_m\partial_n\ln(1+\eta^{-1})=\int d\theta\,\sigma \frac{\dr{q}^k\dr{q}^m\dr{q}^n}{(1+\eta)(1+\eta^{-1})}
\end{equation}
where the subscript $\ell$ is omitted, 
implicitly understanding the 
summation over it. We first use 
the fundamental BA equation
\eqref{basic} obtaining
\begin{equation}
\label{App1e2}
\partial_k \ln\eta = q^k + K*\sigma \partial_k\ln(1+\eta^{-1})\, ,\quad \partial_m\partial_n\ln\eta = K*\sigma \partial_m\partial_n\ln(1+\eta^{-1}) 
\end{equation}
and therefore
\begin{equation}
\label{App1e3}
\int d\theta\, (\underbrace{\partial_k\ln\eta -q^k}_{K*\sigma \partial_k\ln(1+\eta^{-1})})\sigma \partial_m\partial_n\ln(1+\eta^{-1}) = \int d\theta\, \partial_k\ln(1+\eta^{-1})\partial_m\partial_n\ln\eta \, .
\end{equation}
For the l.h.s. of Eq.~\eqref{App1e1} we thus find
\begin{equation}
\label{App1e4}
(A.1)_{\textrm{l.h.s.}}=\int d\theta\, \sigma (\partial_k\ln\eta)\, \partial_m\partial_n\ln(1+\eta^{-1}) - \int d\theta\, \sigma\partial_k\ln(1+\eta^{-1})\partial_m\partial_n\ln\eta \, .
\end{equation}
Finally, we need to calculate the following derivatives
\begin{equation}
\label{App1e5}
\partial_k\ln(1+\eta^{-1})=-\frac{\partial_k\ln\eta}{1+\eta}\, ,\quad \partial_m\partial_n\ln(1+\eta^{-1})
=-\frac{\partial_m\partial_n\ln\eta}{1+\eta} +\frac{(\partial_m\ln\eta)(\partial_n\ln\eta)}{(1+\eta)(1+\eta^{-1})} \, .
\end{equation}
Plugging this into Eq.~\eqref{App1e4} then leads to
\begin{equation}
\label{App1e6}
(A.1)_{\textrm{l.h.s.}}=\int d\theta\, \sigma \frac{(\partial_k\ln\eta)(\partial_m\ln\eta)(\partial_n\ln\eta)}{(1+\eta)(1+\eta^{-1})}=\int d\theta\,\sigma \frac{\dr{q}^k\dr{q}^m\dr{q}^n}{(1+\eta)(1+\eta^{-1})}
\end{equation}
which proves the relation \eqref{JQ_relation}.

\section{Elementary Identity $p_L = 1/n$}
\label{Frac_Id}
In order to prove this identity we need the definitions of the Takahashi-Suzuki (TS)
integers from~\cite{kuniba_continued_1998}, which are collected below.
TS integers are defined for $\gamma/\pi \in Q$
in terms of its continued fraction, which we notate as 
$\gamma/\pi =1/p_0 \equiv [\nu_1, \dots, \nu_\alpha]$ and say has length $\alpha$. As an example take $\gamma = 4 \pi / 9$ 
whose continued fraction will be $\gamma/\pi = 1/p_0= 1/(2 + 1/4 ) \equiv [2,4]$ with length $2$. To make
the notation consistent with the literature on
the Bethe strings we identify our $n$ from $\gamma = n\pi/m$ with the 
TS integer $z_\alpha$ and our $m$ with $y_\alpha$, which coincide with the $\alpha$-th terms of Eq.~\eqref{zInt}
\begin{align}
\label{zInt}
z_\ell = z_{\ell-2} + \nu_{\ell} z_{\ell-1}, && y_\ell = y_{\ell-2} + \nu_{\ell} y_{\ell-1}.
\end{align}
Rational TS numbers $p_\ell$ are obtained in terms of the above integers by
\begin{align}
\label{pInt}
y_\ell = z_\ell p_0 + (-1)^\ell p_{\ell+1},  \text{ with } p_{\alpha+1} =0.
\end{align}

By induction we can show that $p_\alpha = 1/z_\alpha$. The initial induction step for $\alpha=1$ is trivial, as $\gamma =\pi/ \nu_1$ has $p_1 = 1 = 1/1$, which follows from the definitions. 

For our induction hypothesis we take $p_\alpha = 1/z_\alpha$ for a continued 
fraction $\{ \gamma \}\equiv[\nu_1, \dots, \nu_\alpha]$. Consider a 
second anisotropy with continued fraction $\gamma'/\pi=
[\nu_1, \dots, \nu_\alpha, \nu'_{\alpha+1}]$, whose integers 
are denoted $\{z'_\ell, y'_\ell,p'_\ell\}$
with $\ell \in \{0,1, \dots ,\alpha+1\}$. By definition~\eqref{zInt} we know
that the $\gamma$ and $\gamma'$ TS integers $\{z_\ell, y_\ell\}=\{z'_\ell, y'_\ell\}_{\ell < \alpha+1}$
agree for the first $\alpha$ values.

Then beginning from Eq.~\eqref{pInt} with index $i=\alpha$
\begin{align}
y_\alpha &= z_\alpha p'_0 + (-1)^\alpha p'_{\alpha+1},\nonumber \\
y'_{\alpha+1} - y_{\alpha-1}&= \nu'_{\alpha+1} p'_0 z_\alpha + (-1)^{\alpha} p'_{\alpha+1} \nu'_{\alpha+1},\nonumber \\
y'_{\alpha+1} - z_{\alpha-1}+(-1)^\alpha p_\alpha&= \nu'_{\alpha+1} p'_0 z_\alpha + (-1)^{\alpha} p'_{\alpha+1} \nu'_{\alpha+1},\nonumber \\
y'_{\alpha+1} - z_{\alpha-1}p_0 + (-1)^\alpha p_\alpha &= (z'_{\alpha+1} - z_{\alpha-1})p'_0 +(-1)^\alpha \nu'_{\alpha+1} p'_{\alpha+1}.
\end{align}
With $p_0' y_{\alpha+1} = z_{\alpha+1}$ the relation simplifies to 
\begin{align}
z_{\alpha-1}p'_0 - z_{\alpha-1}p_0 + (-1)^\alpha p_\alpha  = (-1)^\alpha \nu'_{\alpha+1} p'_{\alpha+1}.
\end{align}
From the induction hypothesis $p_\alpha=1/z_\alpha$ so obtain
\begin{align}
z_\alpha z_{\alpha-1}p'_0 - z_\alpha z_{\alpha-1}p_0 + (-1)^\alpha   = (-1)^\alpha z_\alpha \nu'_{\alpha+1} p'_{\alpha+1}.
\end{align}
With the relation $p'_0 z_\alpha = y_\alpha - (-1)^\alpha p'_{\alpha+1}$ and Eq.~\eqref{zInt} it follows that
\begin{align}
(-1)^\alpha = (-1)^\alpha z_{\alpha+1} p'_{\alpha+1},
\end{align}
so conclude that $p'_{\alpha+1} = 1/z_{\alpha+1}$ and the identity is proven.

\end{appendix}

\nolinenumbers

\end{document}